\PassOptionsToPackage{english}{babel}
\documentclass[amssymb,amsfonts, superscriptaddress,longbibliography]{revtex4}  

\usepackage[english,ngerman]{babel}
\usepackage{graphicx}  
\usepackage{dcolumn}   
\usepackage{bm}        
\usepackage{amssymb}   
\usepackage{amsmath}
\usepackage{mathtools}
\usepackage[utf8]{inputenc}
 
\usepackage{changes}

\definecolor{myurlcolor}{rgb}{0,0,0.7}
\definecolor{myrefcolor}{rgb}{0.8,0,0}
\usepackage{hyperref}
\hypersetup{colorlinks, linkcolor=myrefcolor,
citecolor=myurlcolor, urlcolor=myurlcolor}

\usepackage{cleveref}
\crefrangeformat{equation}{#3(#1 #4--#5 #2)#6}

\usepackage{comment}
\usepackage{amsthm}

\newcommand{\ignore}[1]{}

\newcommand{\stateB}{\langle\Psi\vert}
\newcommand{\stateK}{\vert\Psi\rangle}

\begin{document}
\selectlanguage{english}

\title{Self-testing Dicke states}

\author{Matteo Fadel}\email{matteo.fadel@unibas.ch} \affiliation{Department of Physics, University of Basel, Klingelbergstrasse 82, 4056 Basel, Switzerland}

\date{\today}

\begin{abstract}

We show that, upon the observation of a specific measurement statistic, Dicke states can be self-tested. Our work is based on a generalization of the protocol considered by Wu \textit{et al.} [PRA \textbf{90} 042339 (2014)], and constitutes a device-independent method for the characterization of a physical device. For realistic situations where experimental imperfections lead to a deviation from the ideal statistics, we give an estimate for the fidelity of the physical state compared to the ideal Dicke state.

\end{abstract}

\maketitle

\section{Introduction}

Self-testing is a process where an untrusted physical realization of states and measurement operations is certified to be equivalent to some ideal reference model.
Our lack of trust requires this certification to be based on a limited number of assumptions, and for this reason it is usually conditioned only on the observed measurement statistics.

The concept of self-testing was first introduced by Mayers and Yao \cite{MY}, who showed how particular statistics originating from a bipartite system can be reproduced only by performing a specific set of local measurements on a maximally entangled pair of qubits. This idea was later generalized to other states also composed by more than two parties, such as the three-qubit $W$ state \cite{stW} and graph states \cite{stGr}.

In this notes we follow the work in \cite{stW} and we generalize the protocol to self test (non trivial) Dicke states, \textit{i.e.} states of $N-k$ qubits in the ground state and $k>1$ qubits in the excited state, symmetric under particle exchange.

\section{Definition of self-testing}

Consider the scenario where $N$ physical devices (observers), labeled by $A_i$ with $i=1...N$, share a $N$-partite state $\stateK$. Every device performs one out of $m_i$ possible local measurements $M_{j_i,A_i}$ (with $j_i=1...m_i$) on its share of the state, and obtains as outcome either $+ 1$ or $-1$. Moreover, the devices cannot communicate to each other during the measurement process. Our task is to certify, without assuming what has been measured, whether the physical realization of such experiment is equivalent to a reference model where the state and the measurements are known.

To be more precise, we formalize this concept by saying that a physical experiment and a reference experiment are equivalent if there exist a local isometry (\textit{i.e.} a map between Hilbert spaces)
\begin{equation}\label{lociso}
\Phi = \Phi_{A_1} \otimes ... \otimes \Phi_{A_N}
\end{equation}
and a state $\vert \text{junk}\rangle$, such that for every $j_i=1, ..., m_i$ and $A_i=1, ..., N$
\begin{align}
\Phi\left( \stateK \right) &= \vert\text{junk}\rangle \otimes \vert\Psi^\star\rangle \label{isoTransfSt}\\
\Phi\left( M_{j_i,A_i} \stateK \right) &= \vert\text{junk}\rangle \otimes M_{j_i,A_i}^\star \vert\Psi^\star\rangle \label{isoTransfMeas}
\end{align}
where $M_{j_i,A_i}^\star$ and $\vert\Psi^\star\rangle$ denote respectively the measurements and state in the reference experiment, and $\vert \text{junk}\rangle$ is in the same Hilbert space as $\stateK$.
This definition of equivalence is motivated by the fact that performing local operations, as well as adding local degrees of freedom (ancillas), do not change the state: one can always exploit the arbitrariness of the reference system and neglect some degrees of freedom to go back to the original state. For this reason, two states mapped one into the other by an isometry are equivalent.

Self-testing consist in the claim that if the correlations observed in a physical experiment coincides with the one predicted by a particular reference experiment, namely
\begin{equation}
\stateB \bigotimes_{i=1}^N M_{j_i,A_i} \stateK = \langle\Psi^\star\vert \bigotimes_{i=1}^N M_{j_i,A_i}^\star \vert\Psi^\star\rangle \;,
\end{equation}
for every choice of measurement settings, then the two experiments are equivalent. This means that physical realizations of states and measurements can be certified to be equivalent to a reference model only by looking at the statistics of the physical measurement outcomes.

For practical purposes, it is important that self-testing protocols require only few measurement settings per party, and that the correlators we want to measure involve only a subset of all possible combinations of measurements.

\section{Self-testing Dicke states}
Dicke state states are symmetric $N$-qubit states with $k$ qubits in $\vert 1 \rangle$ and $N-k$ qubits in $\vert 0\rangle$, in symbols
\begin{equation}
\vert D_N^k \rangle = {{N}\choose{k}}^{-\frac{1}{2}} \text{Sym} \left( \vert 0\rangle^{\otimes N-k} \vert 1 \rangle^{\otimes k} \right) \;,
\end{equation}
where $\text{Sym}(...)$ denotes the symmetrization by particle exchange. The reference experiment we consider is the one where the $N$ qubits of a Dicke state are shared among observers $A_{1}...A_{N}$, each allowed to perform local spin measurements $\sigma_x^{(A_i)}$ or $\sigma_z^{(A_i)}$, except for observer $A_N$ that can additionally measure $(\sigma_x^{(A_N)}+\sigma_z^{(A_N)})/\sqrt{2}$.

Consider now the physical realization of this scenario, where the $N$ observers share a state $\stateK$ and perform local measurements $X_{A_i}$, $Z_{A_i}$ (for $i=1...N$) and $D_{A_N}$. Note that these measurements need not to be spin measurements, and nothing is assumed about them or about the state.

Our claim is that it is possible to conclude that the physical experiment is equivalent to the reference experiment if we observe the statistics 
\begin{equation} \label{statP}
\stateB P_{A_1}^{a_1}...P_{A_N}^{a_N} \stateK  =  {{N}\choose{k}}^{-1} \quad\text{ for all $\vec{a}$ such that }  \left\lVert\vec{a}\right\rVert_1 = k 
\end{equation}
\begin{center}
and
\end{center}
\begin{eqnarray}
\stateB P_{C_1}^{a_1}...P_{C_{N-2}}^{a_{N-2}} X_{C_{N-1}} X_{A_{N}}\stateK & = & 2 {{N}\choose{k}}^{-1} \label{PXX}\\
\stateB P_{C_1}^{a_1}...P_{C_{N-2}}^{a_{N-2}} Z_{C_{N-1}} Z_{A_{N}}\stateK & = & - 2 {{N}\choose{k}}^{-1} \label{PZZ}\\
\stateB P_{C_1}^{a_1}...P_{C_{N-2}}^{a_{N-2}} X_{C_{N-1}} Z_{A_{N}}\stateK & = & 0 \label{PXZ}\\
\stateB P_{C_1}^{a_1}...P_{C_{N-2}}^{a_{N-2}} X_{C_{N-1}} D_{A_{N}}\stateK & = & \sqrt{2} {{N}\choose{k}}^{-1} \label{PXD}\\
\stateB P_{C_1}^{a_1}...P_{C_{N-2}}^{a_{N-2}} Z_{C_{N-1}} D_{A_{N}}\stateK & = & -\sqrt{2} {{N}\choose{k}}^{-1} \label{PZD}
\end{eqnarray}
for all $\vec{a}$ such that $\sum_{i=1}^{N-2} a_i = k-1$ and all $\vec{C}=(C_1,...,C_{N-1})$ cyclic permutations of $(A_1,...,A_{N-1})$, and where $P_{A_{i}}^{a_i} = (1+(-1)^{a_i} Z_{A_i})/2 $ are projectors for the $Z_{A_i}$ measurement.

This claim is the main result of this work, and implies that Dicke states can be self-tested. To prove this statement, we propose an isometry and we use the experimentally observed statistics to show it certifies that the physical experiment is equivalent to the reference experiment. For this proof we will use a number of identities that we will derive in what follows.

\vspace{10mm}
\textbf{Projector identity.} Eq.\eqref{statP} implies
\begin{equation}\label{Proj}
\stateB \sum_{\vec{a}} \delta\left(\left\lVert\vec{a}\right\rVert_1 - k\right) P_{A_1}^{a_1}...P_{A_N}^{a_N} \stateK = 1 \;,
\end{equation}
where $\lVert\vec{a}\rVert_p = \left( \sum_{i=1}^N \vert a_i\vert^p \right)^{1/p}$ is the $p-$norm.
Note that since $\langle\psi\vert\phi\rangle = 1$ if and only if $\vert\psi\rangle = \vert\phi\rangle$, then $\stateB O \stateK = 1$ implies that $\stateK$ is an eigenstate of $O$ with eigenvalue $1$.
Using the fact that $P_{A_i}^{a_i}P_{A_i}^{b_i} = \delta(a_i-b_i) P_{A_i}^{a_i}$ we deduce
\begin{eqnarray}
\stateB \left(\sum_{\vec{a}} \delta\left(\left\lVert\vec{a}\right\rVert_1 - k\right) P_{A_1}^{a_1}...P_{A_N}^{a_N}\right)^2 \stateK &=& \stateB \sum_{\vec{a},\vec{b}} \delta\left(\left\lVert\vec{a}\right\rVert_1 - k\right)\delta(\lVert\vec{b}\rVert_1 - k) P_{A_1}^{a_1}...P_{A_N}^{a_N} P_{A_1}^{b_1}...P_{A_N}^{b_N} \stateK \nonumber\\
&=& \stateB \sum_{\vec{a},\vec{b}} \delta\left(\left\lVert\vec{a}\right\rVert_1 - k\right) \delta(\lVert\vec{b}\rVert_1 - k) \delta(\vec{a}-\vec{b}) P_{A_1}^{a_1}...P_{A_N}^{a_N}\stateK \nonumber\\
&=& \stateB \sum_{\vec{a}} \delta\left(\left\lVert\vec{a}\right\rVert_1 - k\right) P_{A_1}^{a_1}...P_{A_N}^{a_N} \stateK = 1 \;,
\end{eqnarray}
which shows that the operator $\sum_{\vec{a}} \delta\left(\left\lVert\vec{a}\right\rVert_1 - k\right) P_{A_1}^{a_1}...P_{A_N}^{a_N}$ is a projector.

Eq.\eqref{statP} directly implies also
\begin{equation}\label{singleProj}
P_{A_1}^{a_1}...P_{A_N}^{a_N} \stateK  =  \begin{cases} 
      {{N}\choose{k}}^{-\frac{1}{2}}\vert\xi\rangle &\text{ if }  \left\lVert\vec{a}\right\rVert_1 = k \\
      0 &\text{ otherwise} 
   \end{cases}
\end{equation}
where $\vert\xi\rangle$ is some (normalized) state. 

In a similar way we see that for all $\vec{a}$ such that $\sum_{i=1}^{N-2} a_i = k-1$ we have
\begin{equation}\label{eqPPP}
 \stateB (P_{C_1}^{a_1})^2...(P_{C_{N-2}}^{a_{N-2}})^2 \stateK = \stateB (P_{C_1}^{a_1})^2... (P_{C_{N-2}}^{a_{N-2}})^2(P_{C_{N-1}}^{0}+P_{A_{N-1}}^{1}) (P_{A_N}^{0}+P_{A_N}^{1})\stateK =  2 {{N}\choose{k}}^{-1} \;,
\end{equation}
where $\vec{C}=(C_1,...,C_{N-1})$ is a cyclic permutations of $(A_1,...,A_{N-1})$. From Eq.\eqref{eqPPP} we define for later use 
\begin{equation}\label{PHIdef}
P_{C_1}^{a_1}...P_{C_{N-2}}^{a_{N-2}}  \stateK = \sqrt{2 {{N}\choose{k}}^{-1}} \vert\phi\rangle \qquad\Rightarrow\qquad  \vert\phi\rangle = \sqrt{\dfrac{1}{2} {{N}\choose{k}}} P_{C_1}^{a_1}...P_{C_{N-2}}^{a_{N-2}}  \stateK
\end{equation}
where $\vert\phi\rangle$ is some normalized state. 

\vspace{10mm}
\textbf{Operators relabeling identities.} Using the definition of $\vert\phi\rangle$ given in Eq.\eqref{PHIdef}, and the observed measurement statistic Eq.\eqref{PXX}, we obtain
\begin{equation}
\langle \phi \vert X_{C_{N-1}} X_{A_N} \vert \phi \rangle  =  \dfrac{1}{2} {{N}\choose{k}} \stateB P_{C_1}^{a_1}...P_{C_{N-2}}^{a_{N-2}} X_{C_{N-1}} X_{A_N } \stateK  =  1
\end{equation}
which means that $X_{C_{N-1}}\vert \phi \rangle = X_{A_N} \vert \phi \rangle$ or equivalently
\begin{equation}\label{PXPX}
P_{C_1}^{a_1}...P_{C_{N-2}}^{a_{N-2}} X_{C_{N-1}} \stateK  =  P_{C_1}^{a_1}...P_{C_{N-2}}^{a_{N-2}} X_{A_{N}} \stateK \;.
\end{equation}
The statistics in Eq.\eqref{PZZ}, \eqref{PXZ}, \eqref{PXD}  and \eqref{PZD} implies $\langle \phi \vert Z_{C_{N-1}} Z_{A_N} \vert \phi \rangle = -1$, $\langle \phi \vert X_{C_{N-1}} Z_{A_N} \vert \phi \rangle = 0$ and $\langle \phi \vert \left( \frac{X_{C_{N-1}} - Z_{C_{N-1}}}{2} \right) D_{A_N} \vert \phi \rangle = 1$, corresponding respectively to
\begin{equation}\label{PZPZ}
P_{C_1}^{a_1}...P_{C_{N-2}}^{a_{N-2}} Z_{C_{N-1}} \stateK  =  -P_{C_1}^{a_1}...P_{C_{N-2}}^{a_{N-2}} Z_{A_{N}} \stateK \;,
\end{equation}
\begin{equation}
P_{C_1}^{a_1}...P_{C_{N-2}}^{a_{N-2}} X_{C_{N-1}} \stateK  \perp  P_{C_1}^{a_1}...P_{C_{N-2}}^{a_{N-2}} Z_{A_{N}} \stateK \;,
\end{equation}
\begin{equation}\label{PDPD}
P_{C_1}^{a_1}...P_{C_{N-2}}^{a_{N-2}} D_{A_{N}} \stateK =  P_{C_1}^{a_1}...P_{C_{N-2}}^{a_{N-2}}  \left( \dfrac{X_{C_{N-1}} - Z_{C_{N-1}}}{2} \right) \stateK = P_{C_1}^{a_1}...P_{C_{N-2}}^{a_{N-2}}  \left( \dfrac{X_{A_{N}} + Z_{A_{N}}}{2} \right) \stateK \;,
\end{equation}
where for the last equality we used Eq.\eqref{PXPX} and Eq.\eqref{PZPZ}. At this point we would like to emphasize that all the identities derived here are valid for every $\vec{a}$ such that $\sum_{i=1}^{N-2} a_i = k-1$, and for every cyclic permutation $\vec{C}$ of $(A_1,...,A_{N-1})$.

\vspace{10mm}
\textbf{Anticommutation identities.} Using Eq.\eqref{PDPD}, and the fact that $D_{A_N}^2=X_{A_i}^2=Z_{A_i}^2=\mathbb{I}$, we write
\begin{equation}
P_{C_1}^{a_1}...P_{C_{N-2}}^{a_{N-2}} \stateK = P_{C_1}^{a_1}...P_{C_{N-2}}^{a_{N-2}} D_{A_{N}}^2 \stateK = P_{C_1}^{a_1}...P_{C_{N-2}}^{a_{N-2}} \left( \mathbb{I} + X_{A_N} Z_{A_N} + Z_{A_N} X_{A_N} \right) \stateK
\end{equation}
from which is derived the anticommutation relation
\begin{equation}\label{anticomm}
P_{C_1}^{a_1}...P_{C_{N-2}}^{a_{N-2}} X_{A_N} Z_{A_N}  \stateK  = -P_{C_1}^{a_1}...P_{C_{N-2}}^{a_{N-2}} Z_{A_N} X_{A_N}  \stateK \;.
\end{equation}
Again, this identity is valid for every $\vec{a}$ such that $\sum_{i=1}^{N-2} a_i = k-1$, and for every cyclic permutation $\vec{C}$ of $(A_1,...,A_{N-1})$.

\vspace{10mm}
\textbf{Swapping identity.} The relations derived until now allow us to obtain the identity 
\begin{equation}\label{swapping}
P_{A_1}^{a_1} ...P_{A_i}^0...X_{A_j} P_{A_j}^1 ...P_{A_N}^{a_N} \stateK = P_{A_1}^{a_1} ...X_{A_i} P_{A_i}^1... P_{A_j}^0 ...P_{A_N}^{a_N} \stateK \;,
\end{equation}
which is the key tool to prove the main result of this work. To arrive at Eq.\eqref{swapping}, we make use of the fact that
\begin{align}
P_{A_1}^{a_1} ...P_{A_i}^0...X_{A_j} P_{A_j}^1 ...P_{A_N}^{a_N} \stateK &= P_{A_1}^{a_1} ... P_{A_i}^0... P_{A_j}^0 ...P_{A_N}^{a_N} X_{A_j} \stateK &\quad\qquad\text{from Eq.\eqref{anticomm}} \nonumber\\
 &= P_{A_1}^{a_1} ... P_{A_i}^0... P_{A_j}^0 ...P_{A_N}^{a_N} X_{A_i} \stateK &\quad\qquad\text{from Eq.\eqref{PXPX}} \nonumber\\
 &= P_{A_1}^{a_1} ... X_{A_i} P_{A_i}^1... P_{A_j}^0 ...P_{A_N}^{a_N} \stateK &\quad\qquad\text{from Eq.\eqref{anticomm}} \nonumber
\end{align}
which concludes the proof. Note that in the first and in the last step $\sum_{p=1}^N a_p = k$, but we exchanged $a_j \leftrightarrow a_i$.

\vspace{20mm}

\begin{figure}
\includegraphics[width=0.5\columnwidth]{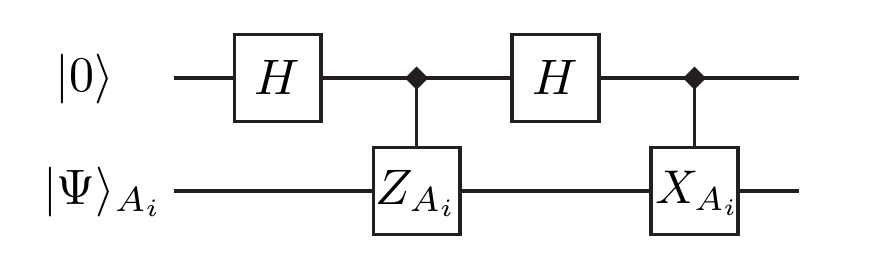}
\caption{\label{localIso} Circuit representing the local isometry for party $A_i$. The upper state is an additional local degree of freedom (ancilla) initially in $\vert 0\rangle$, while the lower state is the partition of $\stateK$ associated with $A_i$. Gate $H$ is the Hadamard transformation.}
\end{figure}

We now have all the tools to show that observing the statistics in Eqs.\eqref{statP}-\eqref{PZD} self-tests the Dicke state $\vert D_N^k \rangle$ and the spin measurements. Indeed, the isometry illustrated in Fig.\ref{localIso} applied on the initial state gives
\begin{align}\label{stState}
\Phi\left( \stateK \vert 0\rangle^{\otimes N} \right) &= \sum_{\vec{a}} X_{A_1}^{a_1}...X_{A_N}^{a_N} P_{A_1}^{a_1}...P_{A_N}^{a_N} \stateK \vert \vec{a} \rangle & \nonumber\\
 &=  \sum_{\left\lVert \vec{a}\right\rVert_1=k} X_{A_1}^{a_1}...X_{A_N}^{a_N} P_{A_1}^{a_1}...P_{A_N}^{a_N} \stateK \vert \vec{a} \rangle & \quad\qquad\text{from Eq.\eqref{singleProj}} \nonumber\\
 &=  \sum_{\left\lVert \vec{a}\right\rVert_1=k} X_{A_1}...X_{A_k} P_{A_1}^{1}...P_{A_k}^{1}P_{A_{k+1}}^{0}...P_{A_N}^{0} \stateK \vert \vec{a} \rangle & \quad\qquad\text{from Eq.\eqref{swapping}} \nonumber\\
 &= X_{A_1}...X_{A_k} P_{A_1}^{1}...P_{A_k}^{1}P_{A_{k+1}}^{0}...P_{A_N}^{0} \stateK \sum_{\left\lVert \vec{a}\right\rVert_1=k}  \vert \vec{a} \rangle & \nonumber \\
 &= \vert \text{junk}\rangle \vert D_N^k \rangle &
\end{align}
where $\vert \text{junk}\rangle$ is some state to be discarded. This concludes the proof that the Dicke state $\vert D_N^k \rangle$ is self-tested.

To see that the measurement operations are also self-tested, we evaluate
\begin{align}\label{stMeasX}
\Phi\left( X_{A_i} \stateK \vert 0\rangle^{\otimes N} \right) &= \sum_{\vec{a}} X_{A_1}^{a_1}...X_{A_N}^{a_N} P_{A_1}^{a_1}...P_{A_N}^{a_N} X_{A_i} \stateK \vert \vec{a} \rangle & \nonumber\\
 &= X_{A_i} \sum_{\vec{a}} X_{A_1}^{a_1}...X_{A_N}^{a_N} P_{A_1}^{a_1}...P_{A_i}^{1-a_i}...P_{A_N}^{a_N}  \stateK \vert \vec{a} \rangle & \quad\qquad\text{from Eq.\eqref{anticomm}} \nonumber\\
 &= \sum_{\vec{a}} X_{A_1}^{a_1}...X_{A_i}^{1-a_i}...X_{A_N}^{a_N} P_{A_1}^{a_1}...P_{A_i}^{1-a_i}...P_{A_N}^{a_N}  \stateK \vert \vec{a} \rangle & \quad\qquad\text{from}\; X_{A_i}X_{A_i}^{a_i} = X_{A_i}^{1-a_i} \nonumber\\
  &= \sum_{\vec{b}} X_{A_1}^{b_1}...X_{A_i}^{b_i}...X_{A_N}^{b_N} P_{A_1}^{b_1}...P_{A_i}^{b_i}...P_{A_N}^{b_N}  \stateK \vert b_1, ..., 1-b_i, ..., b_N \rangle & \quad\qquad\text{def.}\; {\scriptstyle b_l = \delta(l-i)+(-1)^{\delta(l-i)} a_l } \nonumber\\
  &= \sum_{\left\lVert\vec{b}\right\rVert_1=k} X_{A_1}^{b_1}...X_{A_N}^{b_N} P_{A_1}^{b_1}...P_{A_N}^{b_N}  \stateK \vert b_1, ..., 1-b_i, ..., b_N \rangle & \quad\qquad\text{from Eq.\eqref{singleProj}} \nonumber\\
  &=  \sum_{\left\lVert \vec{b}\right\rVert_1=k} X_{A_1}...X_{A_k} P_{A_1}^{1}...P_{A_k}^{1}P_{A_{k+1}}^{0}...P_{A_N}^{0} \stateK \vert b_1, ..., 1-b_i, ..., b_N \rangle & \quad\qquad\text{from Eq.\eqref{swapping}} \nonumber\\
  &= X_{A_1}...X_{A_k} P_{A_1}^{1}...P_{A_k}^{1}P_{A_{k+1}}^{0}...P_{A_N}^{0} \stateK \sum_{\left\lVert \vec{b}\right\rVert_1=k} \vert b_1, ..., 1-b_i, ..., b_N \rangle &\nonumber\\
 &= \vert \text{junk}\rangle \sigma_x^{(A_i)} \vert D_N^k \rangle
\end{align}
which proves that the $X_{A_i}$ measurement act as the Pauli $x$ operator on party $A_i$.

For the $Z_{A_i}$ measurement note that $P_{A_i}^{a_i}Z_{A_i} = (-1)^{a_i} P_{A_i}^{a_i}$, and therefore
\begin{equation}\label{stMeasZ}
\Phi\left( Z_{A_i} \stateK \vert 0\rangle^{\otimes N} \right) = \vert \text{junk}\rangle \sigma_z^{(A_i)} \vert D_N^k \rangle \;,
\end{equation}
proving that the $Z_{A_i}$ measurement act as the Pauli $z$ operator on party $A_i$.

To conclude, the linearity of the isometry allows us to show that
\begin{equation}\label{stMeasD}
\Phi\left( D_{A_N} \stateK \vert 0\rangle^{\otimes N} \right) = \Phi\left( \left(\dfrac{X_{A_N}+Z_{A_N}}{\sqrt{2}}\right) \stateK \vert 0\rangle^{\otimes N} \right) = \vert \text{junk}\rangle \left(\dfrac{\sigma_x^{(A_N)}+\sigma_z^{(A_N)}}{\sqrt{2}}\right) \vert D_N^k \rangle \;.
\end{equation}

Eqs.\eqref{stMeasX}, \eqref{stMeasZ} and \eqref{stMeasD} prove that the experimental measurement operations are also self-tested, meaning that $X_{A_i}$, $Z_{A_i}$ and $D_{A_N}$ are certified to be respectively equivalent to the spin measurements described by $\sigma_x^{(A_i)}$, $\sigma_z^{(A_i)}$ and $(\sigma_x^{(A_D)}+\sigma_z^{(A_N)})/\sqrt{2}$.

\newpage

\section{Robustness}

Inevitable experimental imperfections results in deviations from the ideal measurement statistics. Therefore, we would like to estimate the robustness of the proposed self-testing protocol.
We assume that the discrepancy between the measured and the ideal statistics is at most $\epsilon$, meaning that
\begin{equation}\label{epsilonSim}
\vert \stateB \bigotimes_{i=1}^N M_{j_i,A_i} \stateK - \langle\Psi^\star\vert \bigotimes_{i=1}^N M_{j_i,A_i}^\star \vert\Psi^\star\rangle \vert \leq \epsilon \;,
\end{equation}
for every choice of measurement settings. In this situation we can not conclude that the physical experiment is (exactly) equivalent to the reference experiment, however we can bound tits deviation by saying that
\begin{align}
\left\lVert \Phi\left( \stateK \right) - \vert\text{junk}\rangle \otimes \vert\Psi^\star\rangle \right\rVert_2 \leq \delta \label{StateRob}\\
\left\lVert \Phi\left( M_{j_i,A_i} \stateK \right) - \vert\text{junk}\rangle \otimes M_{j_i,A_i}^\star \vert\Psi^\star\rangle \right\rVert_2 \leq \delta
\end{align}
where $\delta$ is a function of $\epsilon$. This means that physical realizations of states and measurements can be certified to be ``almost'' equivalent to a reference model, with some fidelity dependent on $\delta$.

In what follows we will give an expression for $\delta$, as a function of $\epsilon$, by bounding the norm in Eq.\eqref{StateRob}. Note that in the situation of Eq.\eqref{epsilonSim}, with $\epsilon > 0$, the isometry illustrated in Fig.\ref{localIso} might not be optimal, in the sense that there could be some other isometry giving a better (\textit{i.e.} lower) bound $\delta$, \cite{stW}. Moreover, adding measurements settings might also improve the bound. Here, for simplicity, we will not deal with such optimizations. Considering the same isometry as for the ideal case (Fig.\ref{localIso}), and the same measurement settings, we find a bound for Eq.\eqref{StateRob}.

We define, for compactness, the ideal ``output'' state of the isometry in Fig.\ref{localIso} as (see Eq.\eqref{stState}) $\vert \Theta \rangle = X_{A_1}...X_{A_k} P_{A_1}^{1}...P_{A_k}^{1}P_{A_{k+1}}^{0}...P_{A_N}^{0} \stateK \sum_{\left\lVert \vec{a}\right\rVert_1=k}  \vert \vec{a} \rangle$, and to simplify our calculations we split Eq.\eqref{StateRob} into two terms
\begin{equation}\label{splitFidelity}
\left\lVert \Phi\left( \stateK \right) - \vert\text{junk}\rangle \otimes \vert\Psi^\star\rangle \right\rVert_2  \quad\leq\quad  \left\lVert \Phi\left( \stateK \right) - \vert \Theta \rangle \right\rVert_2 + \left\lVert \vert \Theta \rangle - \vert\text{junk}\rangle \otimes \vert\Psi^\star\rangle \right\rVert_2 \;.
\end{equation}
Here, the first distance is the one between the non-ideal and the ideal output of the isometry considered, and the second distance is the one between the ideal output of the isometry and the ideal target state. In what follows we bound these two terms separately, to obtain an expression for $\delta$ of Eq.\eqref{StateRob}.

\vspace{10mm}
\textbf{First term of Eq.\eqref{splitFidelity}.} Remember that in the previous section we derived from the observed statistics a number of identities involving the measurement operators. Now, observing a deviation from the ideal statistics, Eq.\eqref{epsilonSim}, implies that the identities we derived might still hold approximately, \textit{i.e.} with some error. We start evaluating such errors, to then calculate their effect in the derivation of Eq.\eqref{stState}.

From Eq.\eqref{singleProj} and Eq.\eqref{eqPPP} we estimate
\begin{align}
\left\lVert P_{C_1}^{a_1}...P_{C_{N-2}}^{a_{N-2}} X_{C_{N-1}} \stateK \right\rVert_2 &= \sqrt{\vert\stateB P_{C_1}^{a_1}...P_{C_{N-2}}^{a_{N-2}} X_{C_{N-1}} X_{C_{N-1}} P_{C_{N-2}}^{a_{N-2}} ... P_{C_1}^{a_1} \stateK\vert} \nonumber\\
&= \sqrt{\vert\stateB (P_{C_1}^{a_1})^2...(P_{C_{N-2}}^{a_{N-2}})^2 \stateK\vert} = \sqrt{\vert\stateB P_{C_1}^{a_1}...P_{C_{N-2}}^{a_{N-2}} \stateK\vert} \nonumber\\
& =\sqrt{\vert\stateB P_{C_1}^{a_1}...P_{C_{N-2}}^{a_{N-2}} (P_{C_{N-1}}^{0}+P_{C_{N-1}}^{1}) (P_{A_N}^{0}+P_{A_N}^{1}) \stateK\vert} \nonumber\\
& =\sqrt{ \left| 2{{N}\choose{k}}^{-1} - 4\epsilon \right| } \;,
\end{align}
and, following the same steps
\begin{align}
\left\lVert P_{C_1}^{a_1}...P_{C_{N-2}}^{a_{N-2}} Z_{C_{N-1}} \stateK \right\rVert_2 = \left\lVert P_{C_1}^{a_1}...P_{C_{N-2}}^{a_{N-2}} X_{A_N} \stateK \right\rVert_2 &= \nonumber \\
= \left\lVert P_{C_1}^{a_1}...P_{C_{N-2}}^{a_{N-2}} Z_{A_N} \stateK \right\rVert_2 = \left\lVert P_{C_1}^{a_1}...P_{C_{N-2}}^{a_{N-2}} D_{A_N} \stateK \right\rVert_2 & = \sqrt{ \left| 2{{N}\choose{k}}^{-1} - 4\epsilon \right| } \;.
\end{align}

Now we can estimate the error for Eq.\eqref{PXPX} by computing the norm
\begin{align}
& \left\lVert \left( P_{C_1}^{a_1}...P_{C_{N-2}}^{a_{N-2}} X_{C_{N-1}} - P_{C_1}^{a_1}...P_{C_{N-2}}^{a_{N-2}} X_{A_N} \right) \stateK \right\rVert_2 = \nonumber\\ 
& = \sqrt{\vert \stateB P_{C_1}^{a_1}... X_{C_{N-1}}^2 P_{C_1}^{a_1}... \stateK + \stateB P_{C_1}^{a_1}... X_{A_{N}}^2 P_{C_{1}}^{a_{1}} ... \stateK - 2 \stateB P_{C_1}^{a_1}... X_{C_{N-1}} X_{A_{N}} P_{C_1}^{a_1}... \stateK\vert} \nonumber\\
& = \sqrt{ \left| 2 \left(2 {{N}\choose{k}}^{-1} - 4\epsilon \right) - 2 \left( 2{{N}\choose{k}}^{-1} - \epsilon \right) \right| } = \sqrt{ 6 \vert \epsilon \vert } \;,
\end{align}
and, following the same steps, the error for Eq.\eqref{PZPZ} is
\begin{equation}
\left\lVert \left( P_{C_1}^{a_1}...P_{C_{N-2}}^{a_{N-2}} Z_{C_{N-1}} - P_{C_1}^{a_1}...P_{C_{N-2}}^{a_{N-2}} Z_{A_N} \right) \stateK \right\rVert_2 = \sqrt{ 6 \vert \epsilon \vert}  \;.
\end{equation}

For later use, we bound
\begin{align}
& \vert\stateB P_{C_1}^{a_1}... X_{C_{N-1}} Z_{C_{N-1}} \stateK\vert = \vert\stateB P_{C_1}^{a_1}... X_{C_{N-1}} \left( Z_{C_{N-1}} + Z_{A_N} - Z_{A_N} \right) \stateK\vert \leq \nonumber\\
& \leq \vert\stateB P_{C_1}^{a_1}... X_{C_{N-1}} Z_{A_N} \stateK\vert + \vert\stateB P_{C_1}^{a_1}... X_{C_{N-1}} \left( Z_{C_{N-1}} - Z_{A_N}\right) \stateK\vert = \nonumber\\
& =  \vert\epsilon\vert + \vert\stateB P_{C_1}^{a_1}... X_{C_{N-1}} \left( Z_{C_{N-1}} - Z_{A_N}\right) \stateK\vert \leq \nonumber\\
& \leq \vert\epsilon\vert + \left\lVert P_{C_1}^{a_1}... X_{C_{N-1}}\right\rVert_2  \; \left\lVert P_{C_1}^{a_1}...  \left( Z_{C_{N-1}} - Z_{A_N}\right)\right\rVert_2 = \nonumber\\
& = \vert\epsilon\vert + \left\lVert P_{C_1}^{a_1}... X_{C_{N-1}} \stateK \right\rVert_2  \; \left\lVert P_{C_1}^{a_1}...  \left( Z_{C_{N-1}} - Z_{A_N}\right) \stateK \right\rVert_2 = \nonumber\\
& = \vert\epsilon\vert + \sqrt{ \left| 2 {{N}\choose{k}}^{-1} - 4\epsilon \right|}\sqrt{ 6 \vert \epsilon \vert }  \label{NPXZ}\;
\end{align}
where for the second line we used the triangle inequality $\vert a+b\vert \leq \vert a\vert + \vert b\vert$, and for the fourth the Cauchy-Schwarz inequality $\vert a b\vert \leq \vert a\vert \;  \vert b\vert$.

Eq.\eqref{NPXZ} can now be used to estimate the error for Eq.\eqref{PDPD} as
\begin{align}
& \left\lVert \left( P_{C_1}^{a_1}...D_{A_N} -  P_{C_1}^{a_1}...\dfrac{X_{C_{N-1}} + Z_{C_{N-1}}}{\sqrt{2}} \right)\stateK \right\rVert_2 = \nonumber\\ 
& =  \left( \vert \stateB P_{C_1}^{a_1}...D_{A_N}^2 P_{C_1}^{a_1}... + \dfrac{1}{2}P_{C_1}^{a_1}...X_{C_{N-1}}^2 ... + \dfrac{1}{2}P_{C_1}^{a_1}...Z_{C_{N-1}}^2 ... - P_{C_1}^{a_1}...X_{C_{N-1}}Z_{C_{N-1}} ... - \sqrt{2} P_{C_1}^{a_1}...D_{A_N}X_{C_{N-1}} ... + \right. \nonumber\\
& \phantom{= space} \left.+ \sqrt{2} P_{C_1}^{a_1}...D_{A_N}Z_{C_{N-1}} ... \stateK \vert \right)^{\/2} = \nonumber \\
& =  \sqrt{\left| 2 \left( 2 {{N}\choose{k}}^{-1} - 4\epsilon \right) - 2\sqrt{2}\left( \sqrt{2} {{N}\choose{k}}^{-1} - \epsilon \right) - \stateB P_{C_1}^{a_1}...X_{C_{N-1}}Z_{C_{N-1}} ... \stateK \right|} = \nonumber \\
& =  \sqrt{\vert 2(\sqrt{2}-4)\epsilon +  \stateB P_{C_1}^{a_1}...X_{C_{N-1}}Z_{C_{N-1}} ...  \stateK \vert} \leq \nonumber \\
& \leq  \sqrt{\vert 2(\sqrt{2}-4)\epsilon  \vert + \vert\epsilon\vert + \sqrt{ \left| 2 {{N}\choose{k}}^{-1} - 4\epsilon \right|}\sqrt{ 6 \vert \epsilon \vert }} = \nonumber \\
& = \dfrac{\delta_1}{(2+2\sqrt{2})\sqrt{2}} \;,
\end{align}
where the symbol $\delta_1$ is for compactness, and the denominator is introduced for later convenience.

We now consider the error in commuting $X$ with $Z$.
From the observation that (see also Eqs.(B11) and (B12) of \cite{stW})
\begin{align}
&\left( P_{C_1}^{a_1}... X_{C_{N-1}} Z_{C_{N-1}}  + P_{C_1}^{a_1}... Z_{C_{N-1}} X_{C_{N-1}} \right) \stateK = \nonumber\\
= & \sqrt{2}\left( P_{C_1}^{a_1}... D_{A_N} + \dfrac{P_{C_1}^{a_1}... X_{C_{N-1}} - P_{C_1}^{a_1}... Z_{C_{N-1}}}{\sqrt{2}}\right)\left( P_{C_1}^{a_1}... D_{A_N} - \dfrac{P_{C_1}^{a_1}... X_{C_{N-1}} - P_{C_1}^{a_1}... Z_{C_{N-1}}}{\sqrt{2}}\right) \stateK \;,
\end{align}
we estimate the error in the anticommutator between $X$ and $Z$, Eq.\eqref{anticomm}, by computing (see also Eqs.(B13) and (B14) of \cite{stW})
\begin{align}\label{Nanticomm}
& \left\lVert~\left( P_{C_1}^{a_1}...P_{C_{N-2}}^{a_{N-2}} X_{C_{N-1}} Z_{C_{N-1}}  + P_{C_1}^{a_1}...P_{C_{N-2}}^{a_{N-2}} Z_{C_{N-1}} X_{C_{N-1}} \right) \stateK \right\rVert_2 = \nonumber\\
=& \sqrt{2} \left\lVert  \left( P_{C_1}^{a_1}... D_{A_N} + \dfrac{P_{C_1}^{a_1}... X_{C_{N-1}}}{\sqrt{2}} - \dfrac{P_{C_1}^{a_1}... Z_{C_{N-1}}}{\sqrt{2}} \right)\left( P_{C_1}^{a_1}... D_{A_N} - \dfrac{P_{C_1}^{a_1}... X_{C_{N-1}} - P_{C_1}^{a_1}... Z_{C_{N-1}}}{\sqrt{2}}\right) \stateK \right\rVert_2 \nonumber\\
\leq & \sqrt{2}  \left( \left\lVert P_{C_1}^{a_1}... D_{A_N}\right\rVert\infty + \left\lVert \dfrac{P_{C_1}^{a_1}... X_{C_{N-1}}}{\sqrt{2}} \right\rVert\infty + \left\lVert \dfrac{P_{C_1}^{a_1}... Z_{C_{N-1}}}{\sqrt{2}}\right\rVert\infty \right)\left\lVert \left( P_{C_1}^{a_1}... D_{A_N} - \dfrac{P_{C_1}^{a_1}... X_{C_{N-1}} - P_{C_1}^{a_1}... Z_{C_{N-1}}}{\sqrt{2}}\right) \stateK \right\rVert_2 \nonumber\\
= & \sqrt{2}(2 +2\sqrt{2})\left( \dfrac{\delta_1}{(2+2\sqrt{2})\sqrt{2}} \right)= \delta_1 \;.
\end{align}

Finally, we estimate the error associated to the swapping identity Eq.\eqref{swapping} as
\begin{align}
&\left\lVert \left( P_{A_1}^{a_1} ...P_{A_i}^0...X_{A_j} P_{A_j}^1 ...P_{A_N}^{a_N} -P_{A_1}^{a_1} ... X_{A_i} P_{A_i}^1... P_{A_j}^0 ...P_{A_N}^{a_N}  \right)\stateK \right\rVert_2 = \nonumber\\
& = \left\lVert \left( P_{A_1}^{a_1} ...P_{A_i}^0...X_{A_j} P_{A_j}^1 ...P_{A_N}^{a_N} - P_{A_1}^{a_1} ... P_{A_i}^0 X_{A_i}... P_{A_j}^0 ...P_{A_N}^{a_N} + P_{A_1}^{a_1} ... P_{A_i}^0 X_{A_i}... P_{A_j}^0 - P_{A_1}^{a_1} ... X_{A_i} P_{A_i}^1... P_{A_j}^0 ...P_{A_N}^{a_N}  \right)\stateK \right\rVert_2 \nonumber\\
& = \left\lVert \left( P_{A_1}^{a_1} ...P_{A_i}^0...X_{A_j} P_{A_j}^1 ...P_{A_N}^{a_N} - P_{A_1}^{a_1} ...P_{A_i}^0... P_{A_j}^0 X_{A_j}...P_{A_N}^{a_N} + P_{A_1}^{a_1} ...P_{A_i}^0... P_{A_j}^0 X_{A_j}...P_{A_N}^{a_N} - P_{A_1}^{a_1} ... P_{A_i}^0 X_{A_i}... P_{A_j}^0 ...P_{A_N}^{a_N} + \right.\right.\nonumber\\
&\phantom{= space} \left.\left. + P_{A_1}^{a_1} ... P_{A_i}^0 X_{A_i}... P_{A_j}^0 - P_{A_1}^{a_1} ... X_{A_i} P_{A_i}^1... P_{A_j}^0 ...P_{A_N}^{a_N}  \right)\stateK \right\rVert_2 \nonumber\\
& \leq \left\lVert \left( P_{A_1}^{a_1} ...P_{A_i}^0...X_{A_j} P_{A_j}^1 ...P_{A_N}^{a_N} - P_{A_1}^{a_1} ...P_{A_i}^0... P_{A_j}^0 X_{A_j}...P_{A_N}^{a_N} \right)\stateK \right\rVert_2 + \nonumber\\
&\phantom{+ space} \left\lVert \left(  P_{A_1}^{a_1} ...P_{A_i}^0... P_{A_j}^0 X_{A_j}...P_{A_N}^{a_N} - P_{A_1}^{a_1} ... P_{A_i}^0 X_{A_i}... P_{A_j}^0 ...P_{A_N}^{a_N} \right)\stateK \right\rVert_2 + \nonumber\\
&\phantom{+ space} \left\lVert \left( P_{A_1}^{a_1} ...P_{A_i}^0...X_{A_j} P_{A_j}^1 ...P_{A_N}^{a_N} - P_{A_1}^{a_1} ...P_{A_i}^0... P_{A_j}^0 X_{A_j}...P_{A_N}^{a_N} \right)\stateK \right\rVert_2 \nonumber\\
& = 2\delta_1 + \sqrt{6 \vert\epsilon\vert} \label{errorSwap}\;.
\end{align}

In Eq.\eqref{stState} we used the swapping identity to transform every term of the form $X_{A_1}^{a_1}...X_{A_N}^{a_N} P_{A_1}^{a_1}...P_{A_N}^{a_N} $, with $\left\lVert \vec{a}\right\rVert_1=k$, into $X_{A_1}...X_{A_k} P_{A_1}^{1}...P_{A_k}^{1}P_{A_{k+1}}^{0}...P_{A_N}^{0}$. Note here that every such term is unambiguously defined by the binary vector $\vec{a}$, with $\left\lVert \vec{a}\right\rVert_1=k$. This observation allows us to count how many times we need to apply the swapping identity in Eq.\eqref{stState}, by counting how many ``bit-flips'' are needed to transform one binary vector $\vec{a}$ into the vector $\vec{a}^\star=(1,...,1,0,...,0)$, having $k$ ones followed by $N-k$ zeros.
Recall that the for two strings $s_1$ and $s_2$, the Hamming distance $d_H(s_1,s_2)$ is defined as the number of symbols we need to change to transform $s_1$ into $s_2$. Therefore, a single application of the swapping identity changes a term defined by $\vec{a}$ into an other defined by $\vec{a}^\prime$, with Hamming distance $d_H(\vec{a},\vec{a}^\prime)=2$, since a $0$ is swapped with a $1$. In total, to transform every $\vec{a}$ with $\left\lVert \vec{a}\right\rVert_1=k$ into $\vec{a}^\star$, the number of times we need to apply the swapping identity is
\begin{equation}\label{countSwap}
\sum_{\left\lVert \vec{a}\right\rVert_1=k} \dfrac{d_H(\vec{a},\vec{a}^\star)}{2} =  \frac{\Gamma (n)}{\Gamma (k) \Gamma (n-k)} \;.
\end{equation}
Note that the above number does not depend on which specific $\vec{a}^\star$ we chose among the vectors $\vec{a}$ with $\left\lVert \vec{a}\right\rVert_1=k$.

We have now everything we need to estimate the first term of Eq.\eqref{splitFidelity}. Since the binary vector $\vec{a}$ has $N$ digits, the sum in the first line of Eq.\eqref{stState} involves $2^N$ terms, out of which only ${{N}\choose{k}}$ have $\parallel\vec{a}\parallel_1=k$. We can therefore express
\begin{align}
\left\lVert \Phi\left( \stateK \right) - \vert \Theta \rangle \right\rVert_2 &=  \left\lVert  \sum_{\vec{a}} X_{A_1}^{a_1}...X_{A_N}^{a_N} P_{A_1}^{a_1}...P_{A_N}^{a_N} \stateK \vert \vec{a} \rangle - \vert \Theta \rangle \right\rVert_2 \nonumber\\
&\leq \left\lVert  \sum_{\parallel\vec{a}\parallel_1\neq k} X_{A_1}^{a_1}...X_{A_N}^{a_N} P_{A_1}^{a_1}...P_{A_N}^{a_N} \stateK \vert \vec{a} \rangle \right\rVert_2 + \left\lVert  \sum_{\parallel\vec{a}\parallel_1= k} X_{A_1}^{a_1}...X_{A_N}^{a_N} P_{A_1}^{a_1}...P_{A_N}^{a_N} \stateK \vert \vec{a} \rangle - \vert \Theta \rangle \right\rVert_2 \nonumber\\
&\leq  \sum_{\parallel\vec{a}\parallel_1\neq k}  \left\lVert X_{A_1}^{a_1}...X_{A_N}^{a_N} P_{A_1}^{a_1}...P_{A_N}^{a_N} \stateK \vert \vec{a} \rangle \right\rVert_2 + \left\lVert  \sum_{\parallel\vec{a}\parallel_1= k} X_{A_1}^{a_1}...X_{A_N}^{a_N} P_{A_1}^{a_1}...P_{A_N}^{a_N} \stateK \vert \vec{a} \rangle - \vert \Theta \rangle \right\rVert_2 \nonumber\\
&=  \left( 2^N - {{N}\choose{k}} \right) \vert\epsilon\vert + \left\lVert  \sum_{\parallel\vec{a}\parallel_1= k} X_{A_1}^{a_1}...X_{A_N}^{a_N} P_{A_1}^{a_1}...P_{A_N}^{a_N} \stateK \vert \vec{a} \rangle - \vert \Theta \rangle \right\rVert_2 \nonumber\\
&\leq  \left( 2^N - {{N}\choose{k}} \right) \vert\epsilon\vert + \sum_{\parallel\vec{a}\parallel_1= k} \left\lVert  X_{A_1}^{a_1}...X_{A_N}^{a_N} P_{A_1}^{a_1}...P_{A_N}^{a_N} \stateK \vert \vec{a} \rangle - X_{A_1}...X_{A_k} P_{A_1}^{1}...P_{A_k}^{1}P_{A_{k+1}}^{0}...P_{A_N}^{0} \stateK \vert \vec{a} \rangle \right\rVert_2 \nonumber\\
&\leq \left( 2^N - {{N}\choose{k}} \right) \vert\epsilon\vert + \frac{\Gamma (N)}{\Gamma (k) \Gamma (N-k)} \left( 2\delta_1 + \sqrt{6 \vert\epsilon\vert} \right) \label{BoundFirstTerm} \;,
\end{align}
where in going from the second-to-last to the last row we used Eq.\eqref{countSwap} and Eq.\eqref{errorSwap}.

\vspace{10mm}
\textbf{Second term of Eq.\eqref{splitFidelity}.} To estimate this term, we first find
\begin{align}
\left(\left\lVert \Theta \right\rVert_2 \right)^2& = \left(\left\lVert X_{A_1}...X_{A_k} P_{A_1}^{1}...P_{A_k}^{1}P_{A_{k+1}}^{0}...P_{A_N}^{0} \stateK \sum_{\left\lVert \vec{a}\right\rVert_1=k}  \vert \vec{a} \rangle \right\rVert_2\right)^2 \nonumber\\
& = {{N}\choose{k}} \left(\left\lVert X_{A_1}...X_{A_k} P_{A_1}^{1}...P_{A_k}^{1}P_{A_{k+1}}^{0}...P_{A_N}^{0} \stateK  \right\rVert_2\right)^2 \nonumber\\
& = {{N}\choose{k}} \vert\stateB P_{A_1}^{1}...P_{A_k}^{1}P_{A_{k+1}}^{0}...P_{A_N}^{0} \stateK \vert  \nonumber\\
& = {{N}\choose{k}} \left|  {{N}\choose{k}}^{-1} + \epsilon \right| = \left| 1 + {{N}\choose{k}} \epsilon \right| \;,
\end{align}
which can be used to express
\begin{align}
\left\lVert \vert \Theta \rangle - \vert\text{junk}\rangle \otimes \vert\Psi^\star\rangle \right\rVert_2 & = \left\lVert \vert \Theta \rangle - \dfrac{\vert \Theta \rangle}{\left\lVert \vert \Theta \rangle \right\rVert_2} \right\rVert_2 \nonumber\\
& = \left\lVert \dfrac{\vert \Theta \rangle}{\left\lVert \vert \Theta \rangle \right\rVert_2} \left( \left\lVert \vert \Theta \rangle \right\rVert_2 - 1 \right)  \right\rVert_2 \nonumber\\
& = \left| \sqrt{\left| 1 + {{N}\choose{k}} \epsilon \right|} - 1 \right| \label{BoundSecondTerm} \;.
\end{align}

\vspace{10mm}
\textbf{Bound for Eq.\eqref{splitFidelity}.}
The result Eq.\eqref{BoundFirstTerm}, together with Eq.\eqref{BoundSecondTerm}, is inserted into Eq.\eqref{splitFidelity} to get
\begin{equation}
\left\lVert \Phi\left( \stateK \right) - \vert\text{junk}\rangle \otimes \vert\Psi^\star\rangle \right\rVert_2 \leq \left( 2^N - {{N}\choose{k}} \right) \vert\epsilon\vert + \frac{\Gamma (N)}{\Gamma (k) \Gamma (N-k)} \left( 2\delta_1 + \sqrt{6 \vert\epsilon\vert} \right) + \left| \sqrt{\left | 1 + {{N}\choose{k}} \epsilon \right| } - 1 \right| \;.
\end{equation}
This quantifies how ``close'' the physical state is to the ideal Dicke state.

\section{conclusions}
We have shown that Dicke states can be self-tested. This conclusion is drawn by showing the existence of a local isometry which, together with a specific observed statistic, implies that the physical experiment is equivalent to a reference experiment where specific measurements are performed on a Dicke state. In the practical case where inevitable experimental imperfections cause deviations from the ideal reference statistics, we estimated the robustness of our protocol. We found that even in this case, the physical experiment can still be certified with high fidelity.

\vspace{10mm}
\textbf{Acknowledgements.} I am grateful to Jordi Tura for the useful discussions pointing to the possible extension of Ref.\cite{stW} to Dicke states.

\end{document}